\documentclass[showpacs,superscriptaddress, preprint,aps]{revtex4}

\usepackage{graphicx}
\usepackage{dcolumn}
\usepackage{bm}

\begin{document}

\title{ Low-temperature lattice anomaly in
LaFeAsO$_{0.93}$F$_{0.07}$ probed by x-ray absorption spectroscopy:
Evidence for strong electron-lattice interaction}

\author{C. J. Zhang}
\affiliation{National Institute of Advanced Industrial Science and
Technology, 1-1-1 Umezono, Tsukuba 305-8568, Japan}

\author{H. Oyanagi}
\thanks{To whom correspondence should be addressed. Email address: h.oyanagi@aist.go.jp}
\affiliation{National Institute of Advanced Industrial Science and
Technology, 1-1-1 Umezono, Tsukuba 305-8568, Japan}

\author{Z. H. Sun}
\affiliation{National Institute of Advanced Industrial Science and
Technology, 1-1-1 Umezono, Tsukuba 305-8568, Japan}

\author{Y. Kamihara}
\affiliation{Materials and Structures Laboratory, Tokyo Institute
of Technology, Mail Box R3-1, 4259 Nagatsuta, Midori-ku, Yokohama
226-8503, Japan}

\author{H. Hosono}
\affiliation{Materials and Structures Laboratory, Tokyo Institute
of Technology, Mail Box R3-1, 4259 Nagatsuta, Midori-ku, Yokohama
226-8503, Japan}


\begin{abstract}
The local lattice structure in newly discovered LaFeAsO$_{1-x}$F$_x$
superconductors is studied by extended x-ray absorption fine
structure measurements. An anomalous upturn of the mean-square
relative displacement of the Fe-As bond is detected below $\sim$70 K
as electron carriers are introduced, reflecting the occurrence of
Fe-As bond local lattice fluctuation. Similar to that in cuprates,
this lattice fluctuation exhibits an abrupt depression at the onset
superconducting transition temperature. The results indicate that
strong electron-lattice interaction is involved in the
superconducting transition in oxypcnictide superconductors, putting
a strict limitation on possible theoretical models.
\end{abstract}
 \maketitle

\section{Introduction}
The recently discovered high temperature superconductivity in
fluorine doped LaFeAsO have stirred new interest in the research of
high-$T_c$ superconductors \cite{1}, outside of the cuprate family
\cite{2}. Replacing oxygen by fluorine introduces charge carriers
which is eventually transferred from the La-O(F) charge reservoir
layer to the Fe-As conductive layer. Superconductivity emerges as
the F doping concentration is higher than 5 \%. Subsequent research
suggests that replacing the lanthanum in LaFeAsO with other rare
earth elements such as cerium, samarium, neodymium and praseodymium
leads to superconductors with elevated critical temperature up to 55
K \cite{3,4,5}.

The first and most important question about the LaFeAsO-based new
superconducting system is whether it has similar mechanism for
superconductivity with cuprate superconductors or not. In cuprate
superconductors, the antiferromagnetic properties of the parent
compounds have provoked scenarios of purely electronically driven
superconductivity, where lattice effects are mostly ignored
\cite{6}. On the other hand, various anomalous lattice effects have
recently been observed in cuprates which closely correlate with the
onset of superconducting transition, suggesting that lattice effects
play an important microscopic role in the superconducting pairing
mechanism \cite{7,8,9,10,11,12}. In particular, x-ray absorption
fine structure (EXAFS) measurements have shown that doping causes
local lattice distortion which occurs well above $T_c$
\cite{7,8,11}. Correlating data from inelastic neutron scattering
and inelastic x-ray scattering, isotope effects, Raman spectroscopy,
infrared absorption spectroscopy and femtosecond optical
spectroscopy have been proving that the anomalous local lattice
distortion observed by EXAFS measurements is correlated with the
opening of pseudogap and the formation of polarons \cite{13,14,15}.
The change in dynamics, which is observed across the superconducting
transition temperature, indicates an intimate link of the dynamics
of these polarons with the mechanism of high-temperature
superconductivity \cite{13}. In LaFeAsO-based superconductors,
evidence of pseudogap evolutions similar to the high-$T_c$ cuprates
has been reported \cite{16,17}. However, there is a lack of
experimental data on the lattice effects. In order to find the
appropriate mechanism, lattice effects can provide key information.
In this paper we present results from Fe and As K edge EXAFS
measurements indicating that local Fe-As lattice fluctuation occurs
well above $T_c$. Similar to that in cuprates, this local lattice
fluctuation is closely correlated with the onset of superconducting
transition, indicating that the local lattice fluctuation is
involved in the superconducting coherence in both systems.

\section{Experiment}

Polycrystalline samples LaFeAsO$_{1-x}$F$_x$ ($x$=0; 0.07) were
prepared by solid state synthesis as described elsewhere \cite{1}.
EXAFS measurements were performed at BL13B at Photon Factory,
Tsukuba. Powder samples were mounted on an aluminum holder and
attached to a closed-cycle helium refrigerator. The holder rotates
on a high precision goniometer (Huber 420) to change the incidence
angle. A novel Ge pixel array detector (PAD) with 100 segments was
used in order to gain high throughput and energy resolution. The
detailed description of PAD apparatus was reported elsewhere
\cite{18}. The experimental EXAFS, $\chi$($k$), was analyzed by use
of the Ifeffit analysis package. The fitting to experimental data
was performed in both $R$ space and $k$ space, and the uncertainties
were determined from a reduced $\chi$$^2$ using standard techniques
of error analysis.

\section{Results and discussion}

The Fe and As $K$-edges EXAFS oscillations for LaFeAsO$_{1-x}$F$_x$
($x$=0; 0.07) are measured from 5 K and 300 K. The EXAFS
oscillations at the Fe and As $K$-edges are converted into $k$
space. Typical EXAFS oscillations are shown in Fig. 1(a) for the
$x$=0.07 sample at the Fe $K$ edge (lower panel) and the As $K$ edge
(upper panel). The Fourier transform spectra at 20 K for the Fe and
As $K$ edges are shown as black curves in Fig. 1(b). The position of
the coordination atoms of the Fe and As atoms are also indicated
which are slightly shifted due to the phase-shift effect. The atomic
radial distribution function (RDF) around Fe and As atoms are
simulated using FEFF7. In the simulation the structural parameters
determined by Rietveld analysis are used and all possible scattering
paths (including single-scattering paths and multiple-scattering
paths) are included \cite{19}. The simulated RDF around Fe and As
atoms are shown as the red curves in Fig. 1(b), which can reproduce
all the main peaks in the experimental Fourier transform spectra.

In the EXAFS data analysis process, coordination numbers are set to
the values dictated by the average structure. For the Fe $K$ edge,
we fit the experimental data by including both the nearest
neighboring Fe-As correlation and the next nearest neighboring Fe-Fe
correlation. Typical fitting result is shown as the green curve in
Fig. 1(b). It can be seen that the fitting curve can well reproduce
the experimental data at 1.4$\leq$$R$$\leq$3.1 \AA\ range. For the
As $K$ edge, we fit the experimental data using a single-Gaussian
As-Fe RDF. The result is shown in Fig. 1(b) as the green curve.

Figure 2 gives the temperature dependence of the Fe-As bond
distances and the Fe-Fe bond distances for both samples. It is
obvious that the F-doping leads to a shrink of Fe-As bond distance.
The slight decrease of Fe-As bond distance induced by F doping has
been detected by synchrotron x-ray diffraction measurement
\cite{19}. The contraction of the Fe-As bond distance with F-doping
indicates that the bonding between Fe and As atoms is strengthened
covalent bonding. The contraction of the Fe-As bond distance is
reminiscent of the shortening of the Cu-O bond distance in
La$_{2-x}$Sr$_x$CuO$_4$ as charge carriers are introduced into the
CuO$_2$ plane. In cuprate superconductors, the Cu-O orbital
hybridization is strengthened with the shortening of the Cu-O bond.
According to this fact, we can also suggest that the hybridization
between the Fe 3$d$ orbitals and the As 4$p$ orbitals would be
strengthened. The strengthening of the Fe-As orbital hybridization
favors the flow of charge carrier in the Fe-As conductive layer. It
is also obvious that the Fe-Fe bond distance is shortened in the
F-doped sample, which indicates a decrease of the unit cell volume.
We notice a slight increase of the Fe-Fe bond distance below $\sim$
150 K in undoped LaFeAsO, which is consistent with the tetragonal to
orthorhombic phase transition \cite{19}. In F-doped sample, such a
phase transition disappears. The temperature dependence of Fe-Fe
bond distance for the F-doped sample shows little change in the
whole temperature region.

In Fig. 3 we plot the temperature dependence of the mean-square
relative displacement for the nearest neighboring Fe-As shell
derived from both Fe $K$ edge EXAFS (labeled as
$\sigma$$^2_{Fe-As}$) and As $K$ edge EXAFS (labeled as
$\sigma$$^2_{As-Fe}$) for the LaFeAsO$_{1-x}$F$_x$ ($x$=0.07) sample
together with that of the undoped LaFeAsO sample. As expected, the
results give nearly the same $\sigma$$^2_{Fe-As}$ and
$\sigma$$^2_{As-Fe}$ values at each temperature. At $T$$\geq$150 K
range, the $\sigma$$^2_{Fe-As}$ value decreases with decreasing
temperature for both samples, consistent with the non-correlated
Debye-like behavior. However, below 150 K the temperature dependence
of $\sigma$$^2_{Fe-As}$ exhibits distinctly different behavior. For
undoped LaFeAsO, the $\sigma$$^2_{Fe-As}$ value slightly increases
with further decreasing temperature, which is related to the
so-called spin-density-wave transition \cite{19,20}. For the F-doped
sample, the increase of $\sigma$$^2_{Fe-As}$ at about 150 K is well
suppressed. The $\sigma$$^2_{Fe-As}$ decreases further with
decreasing temperature. Significantly, an anomalous upturn of
$\sigma$$^2_{Fe-As}$ appears at $T$$\leq$70 K. This anomaly occurs
only in F-doped sample while no such anomaly is detected in undoped
parent compound. This anomaly is accompanied by a sharp drop at the
temperature where the onset of superconducting transition occurs
($T_c^{onset}$$\sim$29 K). Similar anomalous behavior was previously
found in La$_{2-x}$Sr$_x$CuO$_4$ samples where an upturn of
$\sigma$$^2_{Cu-O}$ (mean-square relative displacement of the
in-plane Cu-O bond) occurs at $T$$\leq$80 K which is also
accompanied by a sharp decrease at $T_c^{onset}$ \cite{11}. In order
to clearly see the low temperature local lattice instability and its
relation to the $T_c^{onset}$ value, we plot in the inset of Fig. 3
the normalized temperature ($T$/$T_c^{onset}$) dependence of the
mean-square relative displacements for both
LaFeAsO$_{0.93}$F$_{0.07}$ and La$_{1.85}$Sr$_{0.15}$CuO$_4$
samples. It can be seen that a sharp decrease in the mean-square
relative displacement occurs exactly at $T_c^{onset}$ in both
systems. This result indicates that the local lattice instability
might be play an important role in the superconducting coherence in
both systems.

In order to reveal whether or not this anomaly involves the Fe-Fe
bond, we studied the temperature dependence of mean square relative
displacement for the Fe-Fe bond ($\sigma$$^2_{Fe-Fe}$) by fitting
the Fe $K$ edge EXAFS data including the multiple scattering paths.
Figure 4 shows the temperature dependence of $\sigma$$^2_{Fe-Fe}$
for the LaFeAsO$_{1-x}$F$_x$ ($x$=0; 0.07) samples. The temperature
dependence of $\sigma$$^2_{Fe-Fe}$ for the undoped LaFeAsO sample
exhibits a slight increase below $\sim$140 K, which may relate to
the SDW transition \cite{19}. It can be clearly seen that there is
no anomaly in the mean-square relative displacement of the Fe-Fe
bond in $x$=0.07 sample. Thus we conclude that the anomaly in
$\sigma$$^2_{Fe-As}$ below 70 K in F-doped sample involves only the
Fe-As bond. Comparing the mean-square relative displacements of the
Fe-Fe bond in LaFeAsO$_{1-x}$F$_x$, one can find a rather strong
F-doping effect, i.e., the displacement of Fe-Fe bond is strongly
decreased upon F-doping. The temperature dependence of
$\sigma$$^2_{Fe-As}$ shows a complicated behavior related to a
magnetic phase transition. That is, an increase of
$\sigma$$^2_{Fe-As}$ occurs below 140 K in undoped LaFeAsO. In
LaFeAsO$_{0.93}$F$_{0.07}$, the temperature dependence of
$\sigma$$^2_{Fe-As}$ shows no anomaly in the whole temperature
region, consistent with the disappearance of phase transition in
F-doped samples.

To our knowledge macroscopic structural study on LaFeAsO-based
system did not explore any structural transition near 70 K. However,
in cuprate superconductors, a similar upturn of mean-square relative
displacement of the in-plane Cu-O bond has been discovered, which is
related to the splitting of the Cu-O bonds into elongated and
shortened Cu-O bond distances \cite{7}. Based on this fact, we
suggest that a bond splitting of the Fe-As bond in F-doped LaFeAsO
system also occurs below $\sim$70 K. Consequently, some As ions are
shifted forward or backward the adjacent Fe ions. This bond
splitting would lead to a decrease of the magnitude of the Fe-As
(As-Fe) RDF peak. In Fig. 5(a) we plot the Fourier transform
magnitude of the first shell As-Fe bond. In order to compare the
magnitude As-Fe peak quantitatively, we plot the absolute magnitude
of the As-Fe peak in Fig. 5(b). The magnitude of the As-Fe peak
increases with decreasing temperature which is followed by a
decrease below 70 K, consistent with the As-Fe bond splitting model.

In order to quantitatively determine the length scale of the Fe-As
bond-splitting, we plot the Fourier-filtered (back-transforming over
1.4$<R<$2.6 \AA) EXAFS oscillation and amplitude of the first-shell
As-Fe bond at 40 K. The plot is shown in Fig. 5(c). From the EXAFS
oscillation we notice that the local minimum in the amplitude and
the irregularity in the phase near 10.5 \AA$^{-1}$ constitute a
``beat", which signifies the presence of two As-Fe bond distances.
Using the relation $\Delta$$R$=$\pi$/2$k$ between the separation of
the two shells and the position of the beat, the As-Fe distances are
determined to differ by $\sim$0.15 \AA. We notice that the ``beat"
feature is very weak, which possibly comes from two facts: one is
that only small amount of Fe-As bonds are splitted while the other
Fe-As bonds keep undistorted; another reason could be the
unpolarized property of the powder sample used in the EXAFS
measurements.

In doped cuprates, lattice instability is observed as local
distortions (creation of elongated and shortened bonds) probed by
EXAFS which reflects the presence of strong hole-lattice interaction
\cite{7,8,9,10,11}. In case of doped LaFeAsO system, similar
behavior is found in both As and Fe $K$ edge EXAFS. We note that
this distortion is of electronic origin and is different from
crystallographic phase transition as it is observed only after
carrier doping at low temperature. The anomalous change below 70 K
is explained using a local lattice distortion model having equal
number of elongated and shortened Fe-As bonds separated by about
0.15 \AA. Among candidates of distortion models characterized by the
elongated and shortened Fe-As bonds, we consider two cases in Fig. 6
which illustrates the distortion in the FeAs layer in F-doped
LaFeAsO. The left panels show displacement of As atoms (grey ball)
tetrahedrally coordinating with Fe atoms. In the right, four-fold
coordination of Fe atoms is represented by pyramids where each
corner indicates the location of As atom and the direction of
displacement is indicated by arrow. In the upper model all four
Fe-As bonds in the same unit elongate and the shortening of bonds
occur in the adjacent unit, while in the lower model the elongation
and shortening occur in the same unit. In analogy to doped cuprates,
the former and latter models correspond to breezing \cite{21} and
Q$_2$ \cite{22,23} distortions proposed for
La$_{1.85}$Sr$_{0.15}$CuO$_4$, respectively. In the LaFeAsO system,
those two possible distortions may account for the detected
distortion.

We now discuss the implications of the present results. First, the
temperature dependence of mean-square relative displacement of the
Fe-As bond shows remarkable similarity with that of cuprate
superconductors \cite{8,11}. That is, a significant upturn in the
temperature dependence of $\sigma$$^2_{Fe-As}$ in F-doped LaFeAsO
(or $\sigma$$^2_{Cu-O}$ in Sr-doped La$_2$CuO$_4$) occurs at a
characteristic temperature $T^*$, which is related to the opening of
pseudogap in cuprates \cite{9,13,24,25}. In LaFeAsO(F) system, the
opening of pseudogap was recently reported \cite{16,17}, consistent
with the observation of the onset of the upturn of
$\sigma$$^2_{Fe-As}$. We interpret this anomaly as a signature of
lattice instability that indicates the formation of polarons.
Secondly, the increase of mean-square relative displacement
continues until a sudden drop occurs at the onset of superconducting
transition. The plot of mean-square relative displacement \emph{vs}.
normalized temperature ($T/T_c^{onset}$) clearly indicates that the
mean-square relative displacement exhibits a large decrease at the
onset superconducting transition temperature in both Fe-based and
cuprate superconductors, which indicates that the lattice effects
might be important in both systems. However, whether or not the
superconducting mechanism in these systems is driven by
electron-lattice interaction needs further experimental and
theoretical studies.

\section{Conclusion}

In conclusion, we provide evidence from EXAFS measurements that
local lattice instability occurs in F-doped LaFeAsO superconductor,
similar to that in cuprate superconductors. This local lattice
distortion may reveal certain polaron formation well above $T_c$.
The mean-square relative displacements of the Fe-As bond exhibits a
sharp drop at the onset transition temperature, indicating the
lattice effects might be important in this system.

\section{Acknowledgments}

The authors express their greatest thanks to H. Koizumi for
inspiring discussions. The EXAFS experiments were conducted under
the proposal 2007G071 at Photon Factory.

\end{document}